\documentclass[twocolumn,aps,showpacs,preprintnumbers,amsmath,amssymb, superscriptaddress]{revtex4-1}
\pdfoutput=1

\usepackage{bm}
\usepackage{graphicx}
\usepackage{color}

\begin{document}

\title{Electronic band structure of BaCo$_{2}$As$_2$: a fully-doped ferropnictide with reduced electronic correlations}

\author{N. Xu}
\affiliation{Beijing National Laboratory for Condensed Matter Physics, and Institute of Physics, Chinese Academy of Sciences, Beijing 100190, China}
\author{P. Richard}\email{p.richard@iphy.ac.cn}
\affiliation{Beijing National Laboratory for Condensed Matter Physics, and Institute of Physics, Chinese Academy of Sciences, Beijing 100190, China}
\author{A. van Roekeghem}
\affiliation{Beijing National Laboratory for Condensed Matter Physics, and Institute of Physics, Chinese Academy of Sciences, Beijing 100190, China}
\affiliation{Centre de Physique Th\'{e}Žorique, Ecole Polytechnique, CNRS-UMR7644, 91128 Palaiseau, France}
\author{P. Zhang}
\affiliation{Beijing National Laboratory for Condensed Matter Physics, and Institute of Physics, Chinese Academy of Sciences, Beijing 100190, China}
\author{H. Miao}
\affiliation{Beijing National Laboratory for Condensed Matter Physics, and Institute of Physics, Chinese Academy of Sciences, Beijing 100190, China}
\author{W.-L. Zhang}
\affiliation{Beijing National Laboratory for Condensed Matter Physics, and Institute of Physics, Chinese Academy of Sciences, Beijing 100190, China}
\author{T. Qian}
\affiliation{Beijing National Laboratory for Condensed Matter Physics, and Institute of Physics, Chinese Academy of Sciences, Beijing 100190, China}
\author{M. Ferrero}
\affiliation{Centre de Physique Th\'{e}Žorique, Ecole Polytechnique, CNRS-UMR7644, 91128 Palaiseau, France}
\author{A. S. Sefat}
\affiliation{Materials Science and Technology Division, Oak Ridge National Laboratory, Oak Ridge, Tennessee 37831-6114, USA}
\author{S. Biermann}
\affiliation{Centre de Physique Th\'{e}Žorique, Ecole Polytechnique, CNRS-UMR7644, 91128 Palaiseau, France}
\affiliation{Japan Science and Technology Agency, CREST, Kawaguchi 332-0012, Japan}
\author{H. Ding}
\affiliation{Beijing National Laboratory for Condensed Matter Physics, and Institute of Physics, Chinese Academy of Sciences, Beijing 100190, China}

\date{\today}

\begin{abstract}
We report an angle-resolved photoemission spectroscopy investigation of the Fermi surface and electronic band structure of BaCo$_{2}$As$_2$. Although its quasi-nesting-free Fermi surface differs drastically from that of its Fe-pnictide cousins, we show that the BaCo$_{2}$As$_2$ system can be used as an approximation to the bare unoccupied band structure of the related BaFe$_{2-x}$Co$_x$As$_2$ and Ba$_{1-x}$K$_x$Fe$_2$As$_2$ compounds. However, our experimental results, in agreement with dynamical mean field theory calculations, indicate that electronic correlations are much less important in BaCo$_{2}$As$_2$ than in the ferropnictides. Our findings suggest that this effect is due to the increased filling of the electronic 3$d$ shell in the presence of significant Hund's exchange coupling.
\end{abstract}


\maketitle

Although the role they play in ferropnictide superconductivity is still unsettled, there is sufficient evidence indicating that these materials exhibit non-negligible electronic correlations. Hence, previous angle-resolved photoemission spectroscopy studies reported typical overall bandwidth renormalization of about 2-5 \cite{RichardRoPP2011}. Interestingly, Hund's rule coupling was identified as an efficient tuning parameter for electronic correlations \cite{HauleNJP11,AichhornPRB82, LiebschPRB82}, thus indicating the importance of local moments in the physics of these materials. Very recently \cite{WernerNatPhys8}, a study of holed-doped BaFe$_2$As$_2$ including dynamical screening effects evidenced an unusual non-Fermi liquid metallic phase in which partially-screened local moments are frozen above the ordering temperature. Interestingly, the doping-temperature phase diagram of these materials exhibits exotic fractional power-law behaviors of the many-body self-energies assimilated to the phase diagram of  the ``spin-freezing" model \cite{WernerPRL101}, where fractional behaviors occur in some temperature range at fillings close to half-filling. Following this logic, one would expect the strength of the electronic correlations to vary significantly when moving away from the $d^{6-x}$ filling of hole-doped Ba$_{1-x}$K${_x}$Fe$_2$As$_2$, with reduced correlation effects when reaching a $d^7$ filling, thus calling for an experimental characterization of the $d^7$ filling state.  

In this letter, we report ARPES results on BaCo$_2$As$_2$ with a $d^7$ filling. We determine its Fermi surface (FS) and electronic band structure by using polarized photons over a wide energy range. We find that the overall band structure is renormalized by a factor of 1.4 as compared to our local density approximation (LDA) calculations, indicating much weaker correlation effects than in its cousin Ba$_{1-x}$K${_x}$Fe$_2$As$_2$. Nevertheless, we show that BaCo$_2$As$_2$ can be used to approximate the unoccupied states in the 122-ferropnictides. We performed dynamical mean field theory (DMFT) calculations and found good agreement with the experimental ARPES spectra. Our results moreover indicate that the weaker correlation effects in BaCo$_2$As$_2$ as compared to the ferropnictides are mainly a consequence of the larger band filling in the Co-compound. These findings confirm the crucial role played by the Hund's rule coupling in these materials, and allow us to identify the filling of the $d$-band as an efficient tuning parameter for the strength of electronic correlations.

High quality single crystals of BaCo$_2$As$_2$ were grown by the flux method \cite{Sefat_PRB79BaCo2As2}. ARPES experiments were performed at beamlines PGM and Apple-PGM of the Synchrotron Radiation Center (WI) equipped with a Scienta SES 200 analyzer and a Scienta R4000 analyzer, respectively. The energy and angular resolutions were set at 15-30 meV and 0.2$^{\circ}$, respectively. The samples were cleaved \emph{in situ} and measured at 20 K in a vacuum better than 5$\times$10$^{-11}$ Torr. We label the momentum values with respect to the 1 Fe/unit cell Brillouin zone (BZ), and use $c'=c/2$ as the distance between two Fe planes. 

\begin{figure}[!t]
\begin{center}
\includegraphics[width=3.4in]{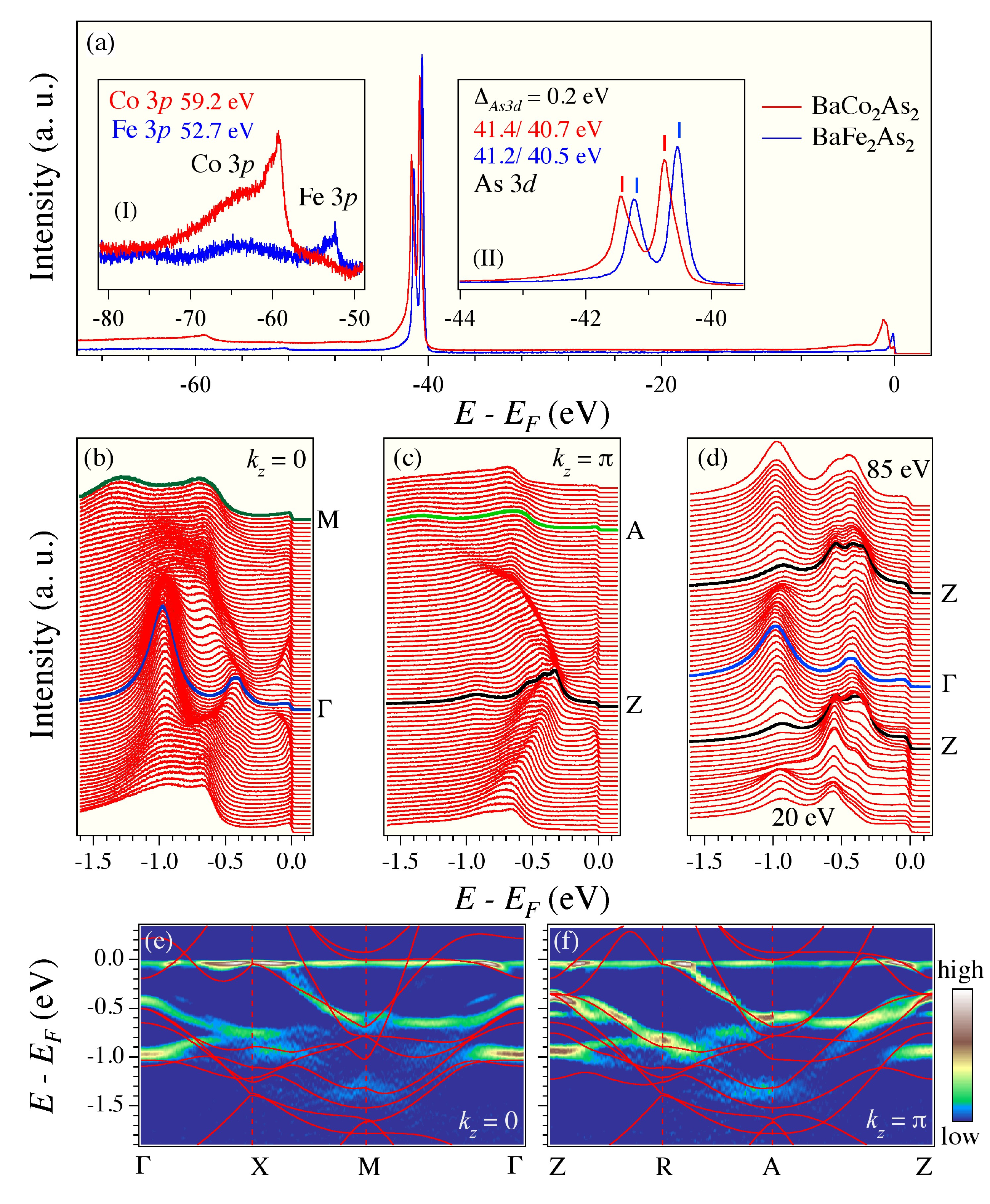}
\end{center}
\caption{\label{Fig1_core_levels}(Color online): (a) Core level spectra of BaCo$_2$As$_2$ (red) and BaFe$_2$As$_2$ (blue) recorded with 195 eV photons. Insets I and II are zooms on the Fe/Co 3$p$ and  As 4$d$ levels, respectively. (b) and (c) EDCs measured at 20 K along $\Gamma$-M ($k_z=0$) and Z-A ($k_z=\pi$), respectively. (d) Photon energy dependence of the EDCs taken at the BZ center. (e) and (f) 2D curvature intensity plots \cite{P_Zhang_RSI2011} along high-symmetry lines in the $k_z=0$ and $k_z=\pi$ planes, respectively. Non-renormalized LDA bands are overlapped for comparison.}
\label{fig1}
\end{figure}

Wide energy photoemission spectra of BaCo$_2$As$_2$ (red) and BaFe$_2$As$_2$ (blue) are compared in Fig. \ref{Fig1_core_levels}(a). A well-defined Co $3p$ core level observed at a binding energy of 59.2 eV in the BaCo$_2$As$_2$ spectrum (inset I) in contrast to 52.7 eV for the Fe $3p$ core level in BaFe$_2$As$_2$ confirms the Fe$\rightarrow$Co substitution. As expected from a previous ARPES study \cite{Neupane_PRB2011} and confirmed by band structure calculations \cite{BerlijnPRL108}, this substitution is accompanied by an upward shift of the chemical potential. Consequently, the As $3d$ core levels (inset II) are observed 200 meV towards higher binding energy in BaCo$_2$As$_2$ as compared with BaFe$_2$As$_2$. We note that the As $3d$ peaks have asymmetric profiles, which is consistent with the enhanced coherence of the BaCo$_2$As$_2$ valence electrons. 

To fully investigate the dispersive states near $E_F$, we performed ARPES experiments over a wide photon energy range. We show energy distribution curves (EDCs) recorded along the $\Gamma (0,0,0)-$M$(\pi,0,0)$ and Z$(0,0,\pi)$-A$(\pi,0,\pi)$ directions in Figs. \ref{Fig1_core_levels}(b) and \ref{Fig1_core_levels}(c), respectively. The two sets of data are different, suggesting a strong 3D character of the electronic structure emphasized by the photon energy dependence of the EDC spectrum at the BZ center displayed in Fig. \ref{Fig1_core_levels}(d). At the first sight, the results are quite different from that of previous studies on the 122-ferropnictides \cite{RichardRoPP2011}. In particular, an electron band rather than hole bands crosses $E_F$ around the $\Gamma$ point. This is indeed what is expected from our LDA calculations, which are shown in Figs. \ref{Fig1_core_levels}(e) and \ref{Fig1_core_levels}(f).  Actually, an overall band renormalization of only 1.4 is needed to capture the essential dispersive features highlighted in the curvature intensity plots \cite{P_Zhang_RSI2011} also shown in Figs. \ref{Fig1_core_levels}(e) and \ref{Fig1_core_levels}(f) for high-symmetry directions. In contrast, the electronic correlations in Ba$_{0.6}$K$_{0.4}$Fe$_2$As$_2$ place the compound at the boundary of an exotic non-Fermi liquid state, where coherent quasi-particles are ill-defined quantities. To fit spectral features to a renormalized LDA band structure requires renormalization of at least a factor 2 \cite{Ding_JPCM2011}. We attribute this decrease in the strength of the electronic correlations to the less drastic effects of Hund's rule coupling \cite{deMedici_PRL107} in the $d^7$ compound BaCo$_2$As$_2$ as compared to the $d^{6-x}$ filling of the $d$-shell in hole-doped BaFe$_2$As$_2$. We come back to this point below.

\begin{figure}[!t]
\begin{center}
\includegraphics[width=3.4in]{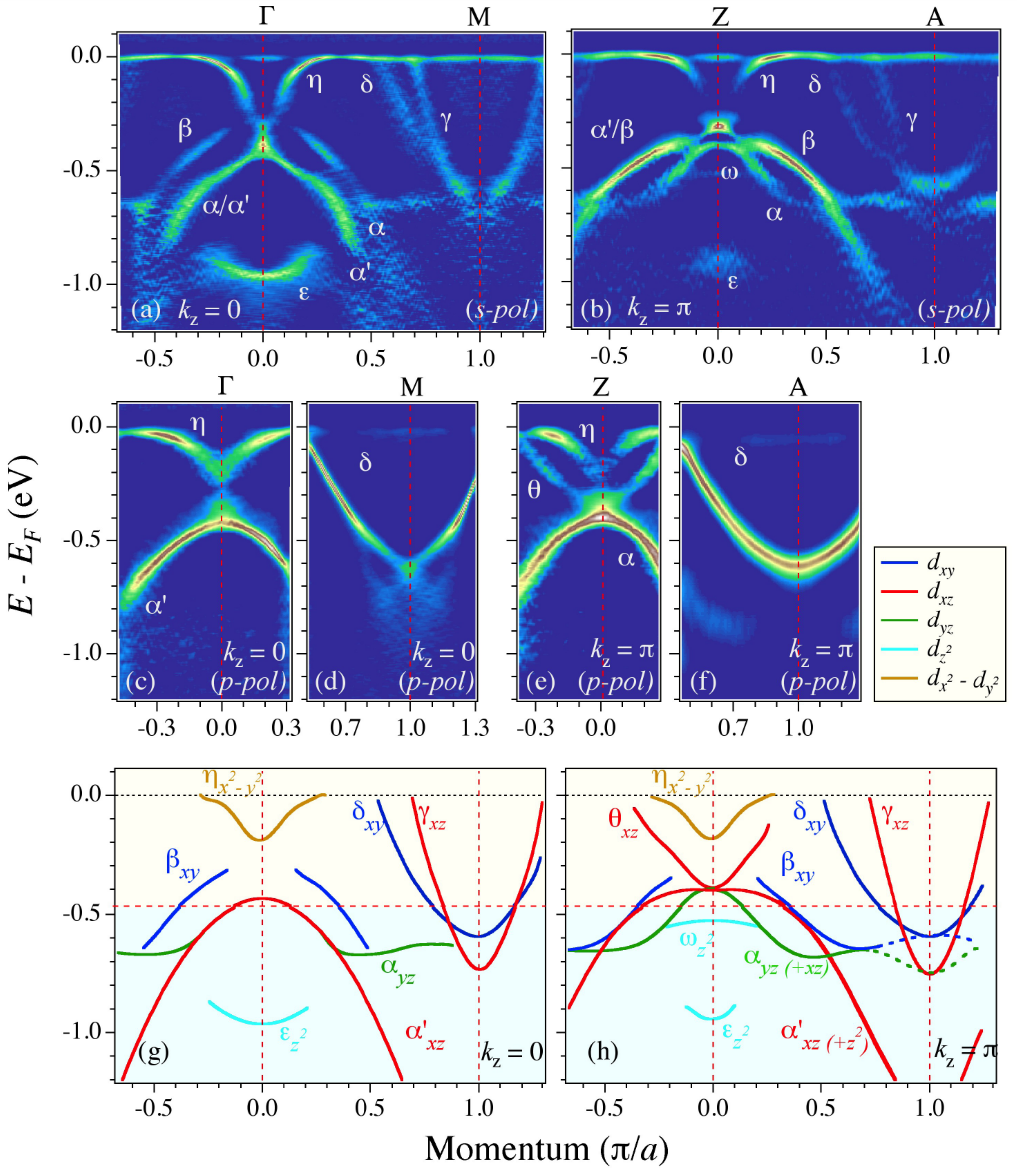}
\end{center}
\caption{\label{Fig2_polarization}(Color online) (a) and (b) 2D curvature intensity plots \cite{P_Zhang_RSI2011} of BaCo$_2$As$_2$ along $\Gamma$-M and Z-A, respectively, recorded in the $s$ polarization geometry. (c)-(f) 2D curvature intensity plots for data recorded with $p$ polarization around (c) $\Gamma$, (d) M, (e) Z and (F) A. (g)-(h) Band dispersions extracted from MDC fits along $\Gamma$-M and Z-A, respectively, along with their orbital characters determined from LDA calculations and the polarization data.}
\label{fig2}
\end{figure}

Photoemission selection rules give further information on the nature of the electronic bands observed. In Figs. \ref{Fig2_polarization}(a) and \ref{Fig2_polarization}(b), we show the curvature intensity plots corresponding to data recorded with a $s$ polarization (with a polarization component along the $z$-axis) along the $\Gamma$-M and Z-A high-symmetry lines, respectively. Similarly, the data recorded with a $p$ polarization are displayed in Figs. \ref{Fig2_polarization}(c) and \ref{Fig2_polarization}(d) for the $k_z=0$ plane and in Figs. \ref{Fig2_polarization}(e) and \ref{Fig2_polarization}(f) for the $k_z=\pi$ plane. While the $s$ polarization enhances bands with odd symmetries with respect to the photoemission plane, the $p$ polarization highlights mainly the even symmetries. Two bands, $\delta$ and $\gamma$, cross $E_F$ at the M(A) point, independently of $k_z$. While both appear clearly with $s$-polarization, only the $\delta$ band is detected with $p$-polarization, suggesting that the other one, the $\gamma$ band, has a odd character. Similarly, two electron bands, called $\eta$ and $\theta$, cross $E_F$ at the BZ center. The first one is detected at $\Gamma$ and Z but appears stronger under $s$ polarization, suggesting an odd character. The effect of the polarization is more severe on the $\theta$ band, which is observed only around the Z point under $p$ polarization. Since the folding potential on the Fe site due to As atoms cannot be neglected \cite{CH_LinPRL107,Brouet_PRB86}, we projected the various orbital characters on our LDA calculated bands in order to facilitate the assignment of the orbital character of each band. Based on our measurements and in comparison with our LDA calculations, we can determine the orbitals characters of all theses bands forming the FS, as well as all the bands down to a binding energy of 1.4 eV. The results are summarized in Figs. \ref{Fig2_polarization}(g) and \ref{Fig2_polarization}(h) for the $k_z=0$ and $k_z=\pi$ planes, respectively.

\begin{figure}[!t]
\begin{center}
\includegraphics[width=3.4in]{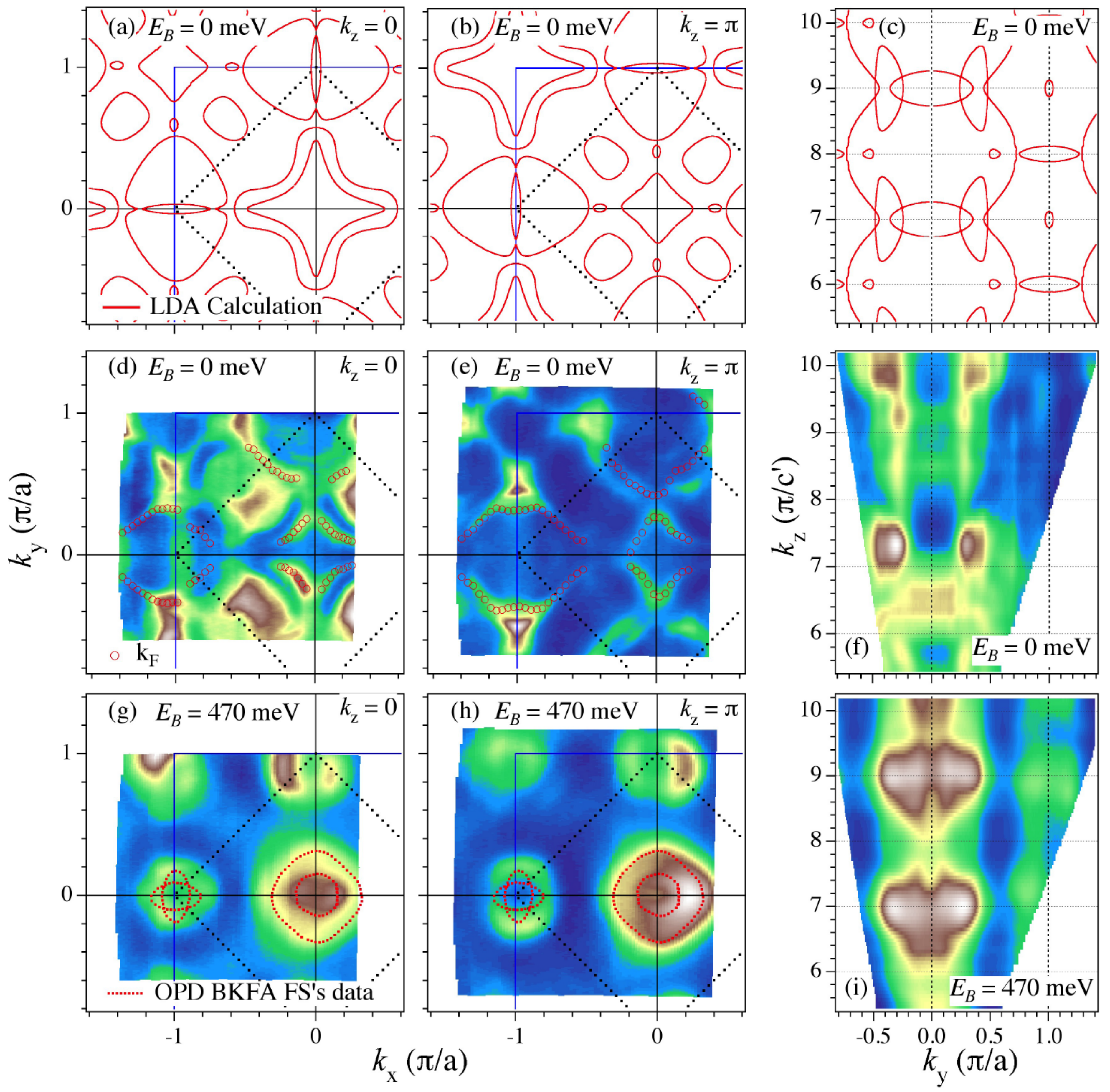}
\end{center}
\caption{\label{Fig_3_comparison}(Color online) (a)-(b) LDA cuts of the FS at $k_z=0$ and $k_z=\pi$,  respectively. (c) LDA FS in the plane defined by $k_y=0$. (d)-(f) Experimental FS mappings corresponding to the cases described in panels (a)-(c). The open symbols represent extracted FS data. (g)-(i) Same as (d)-(f) but for ARPES data recorded 470 meV below $E_F$. The ARPES FS data \cite{Ding_JPCM2011} for Ba$_{0.6}$K$_{0.4}$Fe$_2$As$_2$ (red dashed lines) are plotted in (g) and (h) for comparison.} 
\label{fig4}
\end{figure}

We display in Figs. \ref{Fig_3_comparison}(a) and \ref{Fig_3_comparison}(b) our LDA calculations for the FS cuts at $k_z=0$ and  $k_z=\pi$, respectively. The corresponding ARPES data, obtained by integrating the ARPES intensity within $E_F\pm10$ meV, are shown in Figs. \ref{Fig_3_comparison}(d) and \ref{Fig_3_comparison}(e), respectively. As expected from the previous analysis of the band structure in the vicinity of $E_F$, the FS topology of BaCo$_2$As$_2$ is quite different from that of the Fe-pnictides. For example, the more or less circular hole FS pockets at the $\Gamma$ point of superconducting Ba$_{0.6}$K$_{0.4}$Fe$_2$As$_2$ \cite{Ding_JPCM2011} are replaced by star-shape electron FS pockets in BaCo$_2$As$_2$, thus removing all possibility for electron-hole quasi-nesting scattering in this system. In Fig. \ref{Fig_3_comparison}(f) we show the $k_z$ dispersion at $E_F$ approximated by scanning the photon energy over a wide range, and we display the corresponding LDA calculation in Fig. \ref{Fig_3_comparison}(c). We note that the $\eta$ and $\delta$ bands have a strong 2D character, which is consistent with our assignment of their orbital character. On the other hand, the $\gamma$ band exhibits a small warping that could be caused by a small hybridization with the $p_z$ orbital according to our LDA calculations.

Although the FS of BaCo$_2$As$_2$ differs substantially from that of the 122-ferropnictides, we found out that the constant energy maps recorded at 470 meV below $E_F$ for the $k_z=0$ and $k_z=\pi$ planes, which are respectively displayed in Figs. \ref{Fig_3_comparison}(g) and \ref{Fig_3_comparison}(h), are quite similar to the FSs of Ba$_{0.6}$K$_{0.4}$Fe$_2$As$_2$. To illustrate this resemblance, we overlap on these figures the FSs obtained for Ba$_{0.6}$K$_{0.4}$Fe$_2$As$_2$ in Ref. \cite{Ding_JPCM2011}. In a similar fashion, the $k_xk_z$ map at 470 meV binding energy displayed in Fig. \ref{Fig_3_comparison}(i) could be easily mistaken for the $k_z$ dispersion at $E_F$ of some typical 122-ferropnictides \cite{Malaeb_JPSJ2009, Nan_XuPRB86}.

At least qualitatively, the simplest explanation for this behavior is to assume a $\sim 470$ meV upward shift of the chemical potential following a complete Fe$\rightarrow$Co substitution. As shown in Figs. \ref{Fig2_polarization}(g) and \ref{Fig2_polarization}(h), all the bands in BaCo$_2$As$_2$ below a binding energy of 470 meV find their near-$E_F$ equivalent in the 122-ferropnictides with the same orbital assignment \cite{XP_WangPRB85}. At the $\Gamma$ point for example, the $\alpha$ and $\alpha'$ bands are degenerate over a wide momentum range at $k_z=0$, whereas at $k_z=\pi$ they merge only at the Z point. Their relative position with respect to the $\beta$ band is also the same as in Ba$_{0.6}$K$_{0.4}$Fe$_2$As$_2$ \cite{XP_WangPRB85}. For a more quantitative comparison, we use for the $k_z=0$ data a tight-binding band (TBB) model \cite{KorshunovPRB2008} successfully applied previously to Ba$_{0.6}$K$_{0.4}$Fe$_2$As$_2$ \cite{Ding_JPCM2011}:

\begin{eqnarray}
E^{\alpha, \beta}(k_x, k_y) = & E^{\alpha, \beta}_0+t^{\alpha, \beta}_1(\cos k_x+\cos k_y)\notag\\
& +t^{\alpha, \beta}_2\cos k_x \cos k_y\\
E^{\gamma, \delta}(k_x, k_y) = &E^{\gamma, \delta}_0+t^{\gamma, \delta}_1(\cos k_x+\cos k_y)\notag\\
& +t^{\gamma, \delta}_2 \cos(k_x/2)\cos(k_y/2)
\end{eqnarray}

\begin{figure}[!t]
\begin{center}
\includegraphics[width=3.4in]{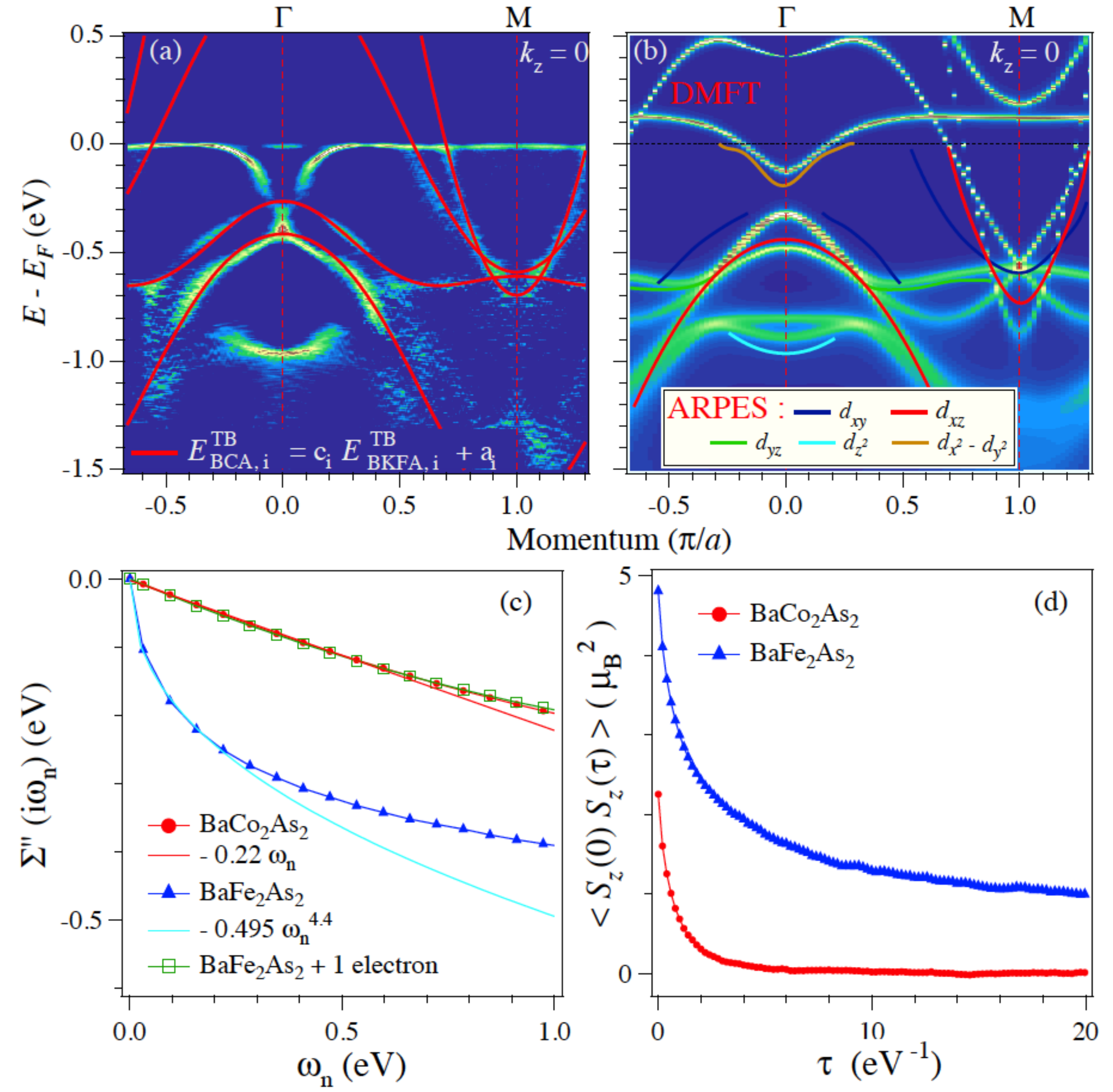}
\end{center}
\caption{\label{Fig4_DMFT}(Color online) (a) 2D curvature intensity plot \cite{P_Zhang_RSI2011} along $\Gamma$-M compared with a TBB fit (red lines). Except for a 470 meV rigid band shift, the bandwidth of the $\beta$ band is four time larger than in Ba$_{0.6}$K$_{0.4}$Fe$_2$As$_2$ \cite{Ding_JPCM2011}, in contrast to two times for all other bands. (b) LDA+DMFT calculation of the band structure of BaCo$_2$As$_2$ (120 K). The experimental bands are overlapped for comparison. (c) Imaginary part of the self-energy (120 K) as a function of the Matsubara frequency for BaCo$_2$As$_2$, BaFe$_2$As$_2$ and the hypothetical compound consisting in BaFe$_2$As$_2$ with one additional electron per Fe. While a linear fit is performed for the BaCo$_2$As$_2$ data, the BaFe$_2$As$_2$ data are fit with a power law $A\omega_n^{\alpha}$, for which we found $\alpha=0.44$. (d) DMFT calculation of the spin-spin correlations in BaCo$_2$As$_2$ and BaFe$_2$As$_2$ at 120 K.}
\end{figure}

\noindent The results of the fit are overlapped in Fig. \ref{Fig4_DMFT}(a) with the curvature intensity plot at $k_z=0$. The agreement with the data is quite good. Disregarding a 470 meV shift for the chemical potential, we find that all the fit parameters, except $t_{1}^{\beta}$ and $t_{2}^{\beta}$ which are 4 times larger, are twice as large as the parameters obtained for Ba$_{0.6}$K$_{0.4}$Fe$_2$As$_2$ \cite{Ding_JPCM2011}. Our experimental data suggest that BaCo$_2$As$_2$ can be viewed in a first rough approximation as a highly overdoped and non-magnetically ordered BaFe$_2$As$_2$ compound, thus allowing us to visualize the unoccupied states in the superconducting 122-ferropnictides. More importantly though, it indicates clearly that BaCo$_2$As$_2$ is less correlated than Ba$_{0.6}$K$_{0.4}$Fe$_2$As$_2$. 

As observed in Ba(Fe$_{1-x}$Ru$_x$)$_2$As$_2$ \cite{Nan_XuPRB86, Brouet_PRL105}, a large decrease in the electronic correlations is expected as 3$d$ electrons are substituted by 4$d$ electrons. This situation is quite different here where the conducting properties of both Fe$^{2+}$ and Co$^{2+}$ derive from 3$d$ electron orbitals with similar spatial extension. To investigate the nature of this unexpected decrease of the electronic correlations in BaCo$_2$As$_2$, we performed LDA+DMFT calculations on this material and on paramagnetic BaFe$_2$As$_2$ using the experimental crystal parameters. The technical details of the calculations are analogous to what was described in ref. \cite{AichhornPRB80}, supplementing the LDA Hamiltonian by explicit local Coulomb interaction terms of density-density form. The Coulomb interaction matrix was parametrized using the standard representation in terms of the Slater integrals $F_0$, $F_2$, $F_4$, which were calculated within the constrained random phase approximation (cRPA) in the implementation of Ref. \cite{Vaugier_PRB2012}. The results given in Fig. \ref{Fig4_DMFT}(c) for BaCo$_2$As$_2$ show a remarkable agreement with the experimental data. We extract a quasi-particle weight $Z$ of 0.8 leading to a bandwidth renormalization of 1.25 that compares well to the experiment. We find a power-law behavior of the self-energy in BaFe$_2$As$_2$ for calculations at the same temperature (120 K), signaling the absence of coherent quasi-particles. However, the corresponding spectral functions are consistent with an apparent renormalisation of the low-energy LDA bands by a factor of 2.9, in agreement with the factor 3 observed experimentally \cite{RichardPRL2010}.

In Fig. \ref{Fig4_DMFT}(c), we plot the imaginary part of the self-energy $\Sigma^{\prime\prime}(i\omega_n)$ of both BaCo$_2$As$_2$ and paramagnetic BaFe$_2$As$_2$ as a function of the Matsubara frequencies. While $\Sigma^{\prime\prime}(i\omega_n)$ varies linearly at low frequency in BaCo$_2$As$_2$, indicating coherent Fermi liquid behavior already at the temperature of the calculation ($T= 120$ K), it follows a nearly square-root behavior characteristic of the ``spin-freezing" regime in BaFe$_2$As$_2$. To investigate further the origin of this phenomenon, we artificially added 1 electron per Fe in BaFe$_2$As$_2$ to mimic the effect of substitution, while maintaining the LDA Hamiltonian of this compound. Surprisingly, the self-energy of this hypothetical crystal can hardly be distinguished from that of BaCo$_2$As$_2$, as shown in Fig. \ref{Fig4_DMFT}(c). This strongly suggests that the decrease in the electronic correlations in BaCo$_2$As$_2$ is mainly due to the less drastic effects of Hund's exchange in the Co-compound whose $d^7$ filling is further away from the case of a half-filled shell. An analogous effect has recently been studied in a three-orbital Hubbard model \cite{WernerPRL101}, where Hund's exchange was found to induce an exotic ``spin-freezing'' regime at finite temperatures for incommensurate fillings close to the half-filled case. This explanation rules out any significant influence from structural variations \footnote{We caution though that structural variations are actually important around a doping level of 6 electrons, especially near the crossover between Fermi liquid and non-Fermi liquid. For instance, trying to hole-dope the BaCo$_2$As$_2$ compound gives a very different picture, with a finite part of $\Sigma^{\prime\prime}(0)$, indicating that this compound is pushed even further into the spin-freezing phase.}. Indeed, this huge change in the level of correlations cannot be linked to a variation of the Coulomb parameter $U$. In both cases, we find $U=F_0$ around 2.8 - 2.9 and $J_H=(F_2+F_4)/14$ around 0.85, with even a small increase of $U$ in the Co case. We furthermore performed additional calculations where we artificially set $J_H$ to 0. In this case, we see a sudden drop of the renormalisation and the suppression of the non-Fermi liquid behavior in BaFe$_2$As$_2$. However, BaCo$_2$As$_2$ only gets slightly less correlated. 

Our results emphasize the importance of local moments in the ferropnictides. In the spin-freezing regime, these unscreened local moments have a long lifetime and spin-spin correlation functions do not decay anymore \cite{WernerPRL101}. As shown by our DMFT calculations of the spin-spin correlations in plotted Fig. \ref{Fig4_DMFT}(d), this situation corresponds to BaFe$_2$As$_2$ at the temperature of the calculations. In contrast, the spin-spin correlations in BaCo$_2$As$_2$ decay rapidly and become vanishingly small in the long-time limit. The apparent strength of the electronic correlations increases with the local moments. Thus, correlations in the ferropnictides can be interpreted as the interaction of local moments with itinerant electrons, as in the Kondo problem. In the present study, we compare two compounds with small orbital polarization (due to the strength of Hund's coupling): BaFe$_2$As$_2$, with six electrons in five bands, and BaCo$_2$As$_2$, which has seven electrons in five bands. In the former case, Hund's exchange drastically reduces the available channels for Kondo screening, thus enhancing the consequences of electronic correlations, while in the latter screening is quite efficient and reestablishes a largely uncorrelated picture. We stress that this effect is not primarily due to the density of states. Indeed, a calculation with 5 degenerate bands gives qualitatively the same conclusion, and the reduction of orbital polarization by Hund's coupling is likely to push the compound close to that model case. Yet, the overlap of $E_F$ with the pseudogap of the $e_g$ orbitals in BaFe$_2$As$_2$ is possibly the cause for orbital-dependent renormalization. This is not the case in BaCo$_2$As$_2$ since $E_F$ locates above the pseudogap, and the $e_g$ orbitals in that case are actually slightly more correlated than the $t_{2g}$ orbitals.

In summary, we used ARPES to characterize the electronic band structure of BaCo$_2$As$_2$. We showed that this material can be used as a first approximation to investigate the unoccupied states of the 122-ferropnictides. However, supported by LDA+DMFT calculations, our study indicates that BaCo$_2$As$_2$ is significantly less correlated than BaFe$_2$As$_2$ due to the Hund's rule coupling and the filling of the $d$ electron shell. Our results reveals clearly the importance of local moments in the physics of the Fe-based superconductors.

\section*{ACKNOWLEDGMENTS}
We acknowledge X. Dai, Z. Fang, Z. Wang and J.-P. Hu for useful discussions, as well as O. Parcollet for help with the numerical calculations. This work was supported by grants from CAS (2010Y1JB6), MOST (2010CB923000 and 2011CBA001000), NSFC (11004232 and 11050110422), computing time at IDRIS/GENCI (under project 1193), the Cai Yuanpei program and the French ANR (project PNICTIDES). This work is based in part upon research conducted at the Synchrotron Radiation Center which is primarily funded by the University of Wisconsin-Madison with supplemental support from facility Users and the University of Wisconsin-Milwaukee. This research was partially supported by the U.S. Department of Energy, Basic Energy Sciences, Materials Sciences and Engineering Division.

\bibliography{biblio_long}

\end{document}